
\input phyzzx
\overfullrule=0pt
\font\boldm=cmbsy10
\def\D{\hbox{\boldm\char'104}}
\def\O{\hbox{\boldm\char'117}}
\def\J{\hbox{\boldm\char'112}}
\font\boldmo=cmr10
\def\oe{\hbox{\boldmo\char'111}}
\font\boldmm=cmfi10
\def\R{\hbox{\boldmm\char'122}}
\def\Z{\hbox{\boldmm\char'132}}
\line{\hfill TAUP-2121-93}
\vskip .1in
\line{\hfill Nov,1993}
\date{}
\vskip 1 true cm
\titlepage
\title{\bf  SEMICLASSICAL vs. EXACT SOLUTIONS OF
CHARGED BLACK HOLE IN FOUR DIMENSIONS  AND EXACT $O(d,d)$ DUALITY}
\vskip 1 true cm
\centerline{\caps ~DAVID ~GERSHON}\footnote{1}{Work supported in
part by the US-Israel Binational
Science Foundation and the Israeli Academy of Sciences.}
\centerline{\it School of Physics and Astronomy}
\centerline{\it Beverly and Raymond Sackler Faculty of Exact Sciences}
\centerline{\it Tel-Aviv University, Tel-Aviv 69978, Israel$\dagger$}
\footnote{\dagger}{e-mail: GERSHON@TAUNIVM.BITNET}
\vskip 2 true cm
\abstract
We derive a charged black hole solution in four dimensions
described by $SL(2,R)\times SU(2)\times U(1)/U(1)^2$ WZW coset model.
Using the algebraic Hamiltonian method we calculate the corresponding
solution that is exact to all orders in ${1\over k}$.
 It is shown that unlike the 2D black
hole, the singularity remains
also in the exact solution, and moreover, in some range of the gauge
parameter the space-time does not fulfil the cosmic censor conjecture,
$\ie$ we find a naked singularity outside the black hole.
Exact dual models are derived as well,  one of them describes a 4D
space-time with
a naked singularity.
Using  the algebraic Hamiltonian approach we also find the exact to all
orders $O(d,d)$
transformation of the metric and the dilaton field for  general
WZW coset models and show the correction
with respect to the transformations in one loop order.
\endpage
\chapter{Introduction}

Since the pioneering paper of Witten on the two dimensional black
hole\Ref\witten{E. Witten\journal Phys.Rev. &D44 (91) 314},
 black holes have vigorously been
studied in string theory\Ref\horowitz{
For a review on stringy black holes
see:  G. T . Horowitz, "The Dark Side of String Theory:
Black Holes and Black Strings", preprint UCSBTH-92-32; "What Is The
True Description of Charged Black Holes?", preprint UCSBTH-92-52}.
Although we do not have an indication whether string theory is the
theory of nature, it is still very helpful to investigate quantum
aspects of black holes in this framework.

Toward a more realistic string theory, we want to investigate black
holes in four dimensions.
All the  vacuum solutions of the Einstein's equations
automatically fulfill the condition for vanishing of the
beta-function to one loop order. However, they are not guarantied
 to correspond to conformal field theories (CFT).
The Schwarzschild, Nordstrom-Reissner  and the Kerr solutions
 are paricular examples.
 We have very limited methods to obtain backgrounds that
correspond to  CFTs: The principal one is to use WZW or WZW coset
models. Up to now,    none of  these black hole
  solutions was  shown to correspond to a WZW coset
model (the Schwarzschild solution
 probably cannot correspond to ungauged WZW
model since the WZW background has an antisymmetric tensor. If it
correspond to gauged WZW, we could at
most hope  to obtain it  after conformal rescaling).
All the    theories that were classified to correspond to 4D
Lorentzian metric\Ref\bers{I. Bars and D. Nemeschanski\journal
Nucl.Phys &B348 (91) 89}
\Ref\gins{P. Ginsparg and F. Quevedo\journal Nucl.Phys. &B385 (92) 527}
 do  not
correspond to any of the above vacuum solutions.
 A classical solution of dilatonic charged black hole in 4D
was derived  in \Ref\garfinkle{D. Garfinkle, G.T. Horowitz and
A. Strominger\journal Phys.Rev.&D43 (91) 3140}
\Ref\gibbons{G.W. Gibbons and K. Maeda\journal Nucl.Phys. &B298 (88)
741}.
It was shown that the presence of the dilaton
field changes the causal structure of the black hole and leads to
curvature singularity at finite radii. The black hole in $\lbrack
\garfinkle\rbrack$ was extensively studied in  connection with
extremal dilatonic black hole (\ie when the inner and the outer
horizons coincide).
It was argued\Ref\preskil{J. Preskil, P. schwartz,
A. Shapere, S. Trivedi and F. Wilczek \journal Mod.Phys.Lett. &A6 (91)
2353}\Ref\holzhey{C.E. Holzhey and F. Wilczek \journal Nucl.Phys.
&B380 (92) 447}
 that such a black hole behaves like an
elementary particle as its spectrum of excitations has an energy gap.
Recently it was shown\Ref\polchinski{S.B. Giddings J. Polchinski and
A. Strominger, "Four Dimensional Black Hole in String Theory",
preprint UCSBTH-93-14}
that at certain limiting cases of the solution  in$\lbrack \garfinkle
\rbrack$ (in which the asymptotic two sphere has a finite radius),
these solutions correspond to
exact string solutions. In these limiting cases
the solutions become a simple  product of the 2D black hole and a non-
singular CFT on the 2-sphere. Other conformal solutions of 4D
charged black holes
were derived in \Ref\david{D. Gershon, "Coset Models Obtained by
Twisting WZW Models and Stringy Charged Black Holes in Four Dimensions",
to appear in Phys.Rev.D}
\Ref\gershon{
D. Gershon, "Exact Solutions of 4D
Black Holes in String Theory", preprint TAUP-2033-92}.

The general procedure to derive sigma models that correspond to
gauged WZW models  is to
parameterize a group and derive its WZW coset models by
integrating out the gauge fields. This procedure is obviously
correct only to one loop order. The exact background has $\O ({1\over
k})$ corrections with respect to the semiclassical solution, where
$k$ is the level.
Up to now, two methods were suggested to derive the exact
to all orders background. The first is the algebraic Hamiltonian
method\Ref\verlinde{R. Dijgraaf, E. Verlinde and H. Verlinde\journal
Nucl.Phys&B371 (92) 269 }
\Ref\bars{I. Bars and K. Sfetsos \journal Phys.Rev. &D46 (92) 4510},
 where one parameterizes the group, writes the Casimir
operators for the zero modes of the Virasoro generators $L_0$,$\bar L_0$
and compares  to the Laplacian in curved space. With this approach
one can derive the exact metric and      dilaton field.
The other method\Ref\tsey{A.A. Tseytlin \journal Nucl.Phys. &B399 (93)
601; preprint CERN-TH.6804/93}\Ref\sfetsos{I. Bars and K. Sfetsos,
\journal  phys.Rev.&D 48 (93) 844)}
is a direct field theoretical approach,
based on replacing the classical WZW action by the exact effective
one and then eliminating the gauge fields, keeping only the local terms.
Both methods coincide, and in the case of  Witten's
 2D black hole yield  a background that was confirmed to satisfy
the beta-function equations  at least
up to the fourth order in $\alpha '$
\Ref\tseyt{A.A. Tseytlin\journal Phys.Lett.&B268 (91)175}\Ref\jack{
I. Jack, D.R.T. Jones and P. Panvel \journal Nucl.Phys.&B393 (93) 95}.
The exact ``black hole'' was shown to have no singularity
\Ref\perry{M. Perry and E. Teo\journal Phys.Rev.Lett. &70 2669 (93)}
although there is an event horizon: In the
exact case a new Eucledian region appears between the singularity and
the black hole interior  and the boundary between the Lorentzian
and the Eucledian regions is a coordinate singularity, which turns
out to be a surface of time reflection symmetry in an extended
space-time.
 One could conjecture that this is how string theory resolves the
problem of space-time singularities in general (and in black holes
in particular),  namely,
     singularities in  one loop order
solutions disappear when introducing all higher orders
 corrections. But this is not the case as we show in this paper.
Another example where this conjecture fails was considered in
\Ref\barssfetsos{I. Bars and K. Sfetsos \journal Phys.Lett. &B301
(93) 183}.
The exact metric that correspond to the semiclassical cosmological
model in \Ref\nappi{C. Nappi and E, Witten
\journal Phys.Lett. &B293 (92) 309} was shown to have a singularity.

The main motivation of this paper is to investigate a solution of
charged black hole in four dimensions and study the higher orders
corrections to the metric.
 Our solution describes an axisymmetric 4D black hole
which is not asymptotically flat, and carries both electric and axionic
charges but no  magnetic charge.
We shall see that the space-time structure in the exact metric
depends strongly on the gauge parameter, unlike in the semiclassical
limit. In all cases the exact metric remains singular, and
moreover, some additional naked singularities can appear.
  In one case we will find that the 4D semiclassical metric   remains
exact to all orders.

The second issue we are interested in is to study duality in the
algebraic Hamiltonian approach and find all the exact dual models
to our black hole solution. We will see that
this approach is natural to
derive the exact $O(d,d)$ transformations
which relate all the dual models. In particular we find one dual
model which describes a naked singularity in space-time (without
a black hole).

$O(d,d)$ symmetries \Ref\veneziano{G. Veneziano\journal Phys.Lett.
&B26(91) 287}\Ref\gasperini{K. Meissner and G. Veneziano\journal
Phys.Lett.&B267 (91) 33}
\Ref\sen{A.Sen\journal Phys.Lett.&B271 (91) 295}
\Ref\hassan{S.F. Hassan and A. Sen\journal Nucl.Phys. &B375 (92) 103}
\Ref\martin{M. Rocek and E. Verlinde \journal Nucl.Phys.&B373 (92)
630}
\Ref\giveon{A.Giveon and M. Rocek\journal Nucl.Phys.&B380 (92)128}
became very popular recently as a helpful tool
to derive semiclassical solutions from known sigma models that
correspond to CFT's.
 These are symmetries of the background that appear
when the background is independent of $d$ of the target space
coordinates. The corresponding duality transformations of the space-time
metric, the antisymmetric tensor and the dilaton field
are known to one loop order.   In this paper we shall
obtain    the  exact to all orders
transformations of the metric and the dilaton field
in general and show   all the higher order corrections to the
semiclassical transformations.\footnote\dagger{In the case of
$SL(2,R)/U(1)$ and $SU(2)/U(1)$ there is a regularization scheme
where the semiclassical background receives no higher order
corrections in $\alpha'$\Ref\tsin{A.A. Tseytlin \journal Phys.Lett.
&B317 (93) 559}. In such a case the semiclassical
$O(d,d)$ symmetry transformations are  also exact
\Ref\amit{A. Giveon and E. Kiritsis, "Axial-Vector Duality as a Gauged
Symmetry and Topology Change in String Theory",
 preprint CERN-TH.6816/93}.}

The paper is  organized as follows:
In section 2 we derive the charged black hole solution as an
$SL(2,R)\times SU(2)\times U(1)/U(1)^2$ WZW coset model. This is
obtained  by integrating out the gauge fields in the gauged sigma
model action.  In section 3 we derive the exact solution that
correspond to the semiclassical solution of section 2 and analyse it
with respect to the semiclassical limit. Here we show that the
exact metric remains singular. Moreover, the space-time described
by the exact metric depends drastically on the relation between
the  gauge parameter and the levels. In some range of the gauge
parameter the exact solution
 describes a charged black hole, while in some other ranges
 naked singularities  appear on  cone surfaces or on an infinite
string, which cross  the event horizon,  so strictly
speaking the solution does not describe a black hole and
 the space-time becomes non-physical.
In section 4 we derive an expression for the metric, the space-time
gauge fields and the dilaton field for all the dual models. These
are related by the {\it exact} $O(3,3)$ symmetry. In
particular  we derive a dual model which describes a naked singularity
in 4D spacetime. In section 5 we obtain the exact
abelian $O(d,d)$ symmetry transformations of the metric and the dilaton
field for general WZW coset models.  We show that when writing
the inverse exact metric as composed of the one loop order part plus
the $\O({1\over k})$ corrections, the former part transforms
exactly as in  the one loop order $O(d)\times O(d)$ transformations
while the latter part is unchanged. Therefore, knowing  both
the antisymmetric tensor to one loop order and the exact
metric is enough to obtain all the exact dual models.
  Section 6 is reserved for summary and discussion.

\chapter {Four Dimensional Charged Black Hole Solution}

In this section we shall construct  CFTs derived from
$SL(2,R)_{k_1}\times SU(2)_{k_2}\times U(1)/U(1)^2$ WZW coset model which
describe charged black hole in four dimensions in the closed bosonic
string theory. The model we describe here is based on our previous works
$\lbrack \gershon,\david\rbrack$.\footnote{\dagger}{In \Ref\horava{P.
Horava  \journal Phys.Lett.&B 278 (92) 101}
and in $\lbrack\nappi\rbrack$ related coset models were derived,
leading to other 4D cosmological solutions.}
It can be embedded also in
the framework of superstring theories or the heterotic strings
(\ie starting with $N=1$ or $N={1\over 2}$ supersymmetric WZW).
We shall concentrate here only in the closed bosonic strings.
The   central charge of our model is
$c={3k_1\over k_1-2}+{3k_2\over
k_2+2}-1$.
In order that this model describes
the complete space-time we need to have either
$c=26$ in the bosonic strings or $c=15$ in the $N=1$ superstrings.
Alternatively  we can  describe our space time as a tensor product
$M^4\times K$ where $M$ is the four dimensional Lorentzian
space-time and $K$ is some internal space, represented by
another CFT
\Ref\green{ M. Green, J. Schwartz and E. Witten, Superstrings
 Theory 2, Cambridge
    University press (1986)}
so that the total central charge is 26 (or 15 in the supersymmetric
case).
 In this case our model can also be regarded as a Kaluza
 Klein model, with one compactified dimension that is part of $K$.
In both pictures, for  any integer $k_2$ we can find the
appropriate $k_1$.

In this section we shall obtain the
background by integrating out the gauge fields.
Therefore the model is correct to one loop order (a semiclassical
solution). It is clear that this model will have
corrections of order  $\O({1\over k_1}),\O({1\over k_2})$
in the space-time metric, the antisymmetric tensor,
the gauge fields and in the dilaton field. A priory,
only when $k_1,k_2\rightarrow
\infty$ this model can be regarded as  exact to all orders.
(In the $N=1$ supersymmetric case we expect it to be exact to all orders
in ${1\over k}$ also for finite $k$, based on $\lbrack \jack\rbrack$).

 To describe closed bosonic strings which have space-time
 gauge fields in their massless spectrum
\Ref\gross{D.J. Gross, J.A. Harvey, E. Martinec and J.Perry,
Nucl.Phys.Lett.{\bf 54}, 502 (1985)}
\Ref\callan{ C.G. Callan, D. Friedan, E.J. Martinec and M.J. Perry,
Nucl. Phys.{\bf B262}, 593, (1985)}\Ref\ishibashi{ N. Ishibashi,
 M. Li and A. Steif\journal Phys.Rev.Lett. &67 (91) 3336} we use
the fields $X^{\mu}$ which are the space time coordinates (in
our case $\mu=0,...,3$)
 and compactified free bosonic fields $X^a$ which realizes the Kac-Moody
currents of the gauge group $\tilde G$, with $a=1,...,\dim \tilde G$.
 In our model we seek $U(1)$ space-time
gauge fields, thus have  one  compactified field
which we denote by $Z$.
The sigma model action which we will derived correspond to
$$S={1\over 2\pi} \int d^2\sigma
 (G_{\mu\nu}(X) +B_{\mu\nu}(X))\partial _+X^{\mu}
\partial_-X^{\nu} +\partial_-Z\partial_+ Z $$ $$+
A_{\mu}(X)\partial
_+X^{\mu}\partial_-Z
-{1\over 8\pi}\int d^2 \sqrt{h}R^{(2)}\Phi(X)  \eqn\bgop$$
where $G_{\mu\nu}$ is the space-time metric, $B_{\mu\nu}$ is
the antisymmetric tensor, $A_{\mu}$  is the
background space-time gauge field (the electromagnetic vector),
$h$ is the determinant of the world sheet metric,
$R^{(2)}$ is the curvature of the worldsheet and
$\Phi$ is the dilaton field. (Notice that in the heterotic strings
we fermionize the bosonic field $Z$ which will
contribute  only to the right
moving sector). The $U(1)$ symmetry transformation
that corresponds to this action
is $$\delta Z(\sigma)=f(X(\sigma))$$
$$\delta A_{\mu}(X)=-2\partial_{\mu}f(X(\sigma))$$
$$\delta G_{\mu\nu}=\delta B_{\mu\nu}=A_{\mu}(X)\partial_{\nu}f(X)
\eqn\skjhs$$with $f$ an infinitesimal function.

The WZW action
\Ref\wess{ J.~Wess and B. Zumino, Phys.Lett.{\bf B37} (1971)}
\Ref\bosonization{E. Witten, Comm.Math.Phys.
92  (1984) 455}
for a group $G$ is
$$S_0(g)={-k\over 4\pi}\int_{\Sigma}d^2\sigma
Tr(g^{-1}\partial_+ g g^{-1}
\partial_- g) -\Gamma  \eqn\one $$  where  $g$ is an
element of the group
$G$ and $\Gamma$ is the Wess Zumino
term $$\Gamma={ik\over 12\pi}\int_B Tr(g^{-1}dg\wedge
g^{-1}dg \wedge
g^{-1}dg) \eqn\two$$    $B$ is the manifold whose boundary
 is a Riemann surface $\Sigma$.
We use Lorentzian metric on the worldsheet $\Sigma$.

Consider the group $G=SL(2,R)\times SU(2)\times U(1)$.
We denote the group elements $g$ by the direct product
$(h_1,h_2, e^{iX})$,
where    $h_1\in SL(2,R)$,    $h_2\in SU(2)$ and $X$ is a free
compactified $U(1)$ field (\ie $X\sim X+2\pi R$, where $R$ is the
radius of compactification).
The ungauged action is
$S_0(g)=S_0(h_1)+S_0(h_2)+S_0(X)$.
We denote the level of the $SL(2,R)$ WZW by $k_1$ and that  of
$SU(2)$  by $k_2$.
In order that the action is uniquely defined
the level of the compact group $SU(2)$ should be integer
$\lbrack 5\rbrack$.
   Now we  parameterize the group elements of
   $SL(2,R)$   and $SU(2)$
by $$h_1=\exp({t_L\over 2}\sigma_3)\exp (r\sigma_1)
\exp({t_R\over 2}\sigma_3)$$
 $$h_2=\exp(i{\phi_L\over 2}\sigma_3)\exp (i\theta\sigma_1)
\exp(i{\phi_R\over 2}\sigma_3)\eqn\three$$
where $\sigma_i$ are the Pauli matrices.
Expressed in terms of these coordinates, the ungauged action is
$$S_0={k_1\over 8\pi}\int d^2\sigma(4\partial_+r\partial_-r
+\partial_+t_L\partial_-t_L+\partial_+t_R\partial_-t_R+
2\cosh 2r\partial_+t_R\partial_-t_L)$$
$$+{k_2\over 8\pi}\int d^2\sigma(4\partial_+\theta\partial_-\theta
+\partial_+\phi_L\partial_-\phi_L+\partial_+\phi_R\partial_-\phi_R+
2\cos 2\theta\partial_+\phi_R\partial_-\phi_L)$$
$$+ {1\over 4\pi}\int d^2\sigma \partial_+X\partial_-X\eqn\four$$
and   $k_1,k_2$ are taken to be positive.
To obtain the $SL(2,R)$ WZW we have
substituted $h_1$ and took $-k_1$, since we want the $SL(2,R)$
manifold to have  signature $(-++)$ and not $(+--)$.
Next we wish to gauge a diagonal $U(1)^2$ subgroup.
The standard way to gauge  $H_L\times H_R$ subgroup of
$G_L\times G_R$ in
WZW action
\Ref\goddard{ P.~Goddard, A. Kent and D. Olive, Phys.Lett.{\bf B152}
 (1985) 88; W. Nahm, Duke Math.J. ,54  (1987) 579;
E. Guadagnini, M. Martellini and M. Minchev, Phys.Lett.{\bf  191}
 (1987) 69; K. Bardakci, E. Rabinovici, B. Saring, Nucl.Phys.
{\bf  B299} (1988) 157;
D. Altshuster, K.Bardakci, E. Rabinivici, Comm.Math. Phys.
{\bf  118}
241; K. Gawedzki, A. Kupiainen, Phys. Lett. {\bf  B215} (1988) 119;
Nucl. Phys.
{\bf  B320}  (1989) 625;
D. Karabali, Q. Park, H. Schnitzer and Y. Yang, Phys.Lett.
{\bf  B216}
  (1989) 307;  H. Schnitzer, Nucl. Phys. {\bf  324}   (1989) 412;
D. Karabali and H. Schnitzer, Nucl.Phys.{\bf  B329}   (1990) 649}
is by replacing derivatives by covariant derivatives.
The gauged action is
$$S(A,B,g)=S_0(g)+{k\over 2\pi}\int d^2\sigma
\tr (A_-g^{-1}\partial_+g -B_+\partial_-gg^{-1} +B_+gA_-g^{-1})
$$$$-{k\over 4\pi}\int d^2\sigma \tr (A_+A_-+B_+B_-)\eqn\five$$
where  the symmetry
transformation is  $\delta g=vg-gu,\;\;\delta A_i=-D_i u,
\;\;\delta B_i=-D_i v$. \footnote{\dagger}{In the axial gauging it
is more common to transform $A\rightarrow -A$ so that $A,B$ have the
same gauge transformation.}
However the WZ term $\Gamma (g)$ has
a gauge invariant extension only if one restricts to an ``anomaly-
free" subgroup of $G_L\times G_R$.
Denote the generators of $H_R$ and $H_L$ by
$T_{a,L}$ and $T_{a,R}$.  The anomaly free condition is the following
\Ref\holomorphic{E. Witten \journal Commun.Math.Phys. &144 (92)189}:
 $$\tr T_{a,L}T_{b,L}=\tr T_{a,R}T_{b,R}
\;\;\;\;{\rm for}\; a,b =1...{\rm dim}H\eqn\six$$
($\tr$ is the trace on the $G_L\times G_R$ Lie algebra. When
 $G$ is a product of groups $G_i$  with levels $k_i$ this reads
$\tr=\Sigma k_i\tr_i$ where $\tr_i$ is  the trace  in the
representations of the Lie algebra of the group $G_i$).
We shall gauge a (axial) $U(1)^2$  subgroup of
 $SL(2,R)\times SU(2)\times U(1)$,
 generated by   the following
infinitesimal gauge transformations:
$$\delta h_1=
(\epsilon_1\sin\psi\sin\alpha+\epsilon_2\cos\alpha)
{\sigma_3\over 2} h_1+h_1
{\sigma_3\over 2}(\epsilon_1\sin\varphi\sin\beta+\epsilon_2
\cos\beta) $$
$$\delta h_2=\sqrt{k_1\over k_2}
(\epsilon_1\sin\psi\cos\alpha-\epsilon_2\sin\alpha)
{i\sigma_3\over 2} h_2+\sqrt{k_1\over k_2}h_2{i\sigma_3\over 2}
(-\epsilon_1\sin\varphi\cos\beta+\epsilon_2
\sin\beta) $$  $$\delta e^{iX}=i\sqrt{k_1\over 2}\epsilon_1(\cos\psi
e^{iX}+e^{iX}\cos\varphi)\eqn\uyfl$$
where  $\epsilon_1$, $\epsilon_2$ are infinitesimal
 and $\alpha,\beta , \psi,\varphi$  are arbitrary.
(Notice that since in the $SL(2,R)$ WZW we had to take $-k_1$ for
$h_1$, here the anomaly free condition is
$-k_1\tr_{SL(2,R}+k_2\tr_{SU(2)}+\tr_{U(1)}$.)
  To gauge the above symmetry we introduce two abelian gauge fields
 $A_1,A_2$ that  transform as
$$\delta A_{1,i}=-\partial_i\epsilon_1\;\;\;;\;\;\;\;
\delta A_{2,i}=-\partial_i\epsilon_2  \eqn\eight$$
The full gauged  action is the following:
$$S(A_1,A_2,g)=S_0(g)
+{k_1\over 4\pi}\int d^2\sigma
(\sin\psi\sin\alpha {A_1}_++\cos\alpha {A_2}_+)(\partial_-t_R+\cosh 2r
\partial_-t_L)$$$$ +(\sin\varphi\sin\beta {A_1}_-+\cos\beta
{A_2}_-)(\partial_+t_L+\cosh 2r\partial_+t_R)$$  $$
+{\sqrt{k_1 k_2}\over 4\pi}
\int d^2\sigma (\sin\psi\cos\alpha
{A_1}_+-\sin\alpha {A_2}_+)(\partial_-\phi_R+\cos 2\theta\partial_-
\phi_L)$$$$+(-\sin\varphi\cos\beta{A_1}_-+\sin\beta {A_2}_-)
(\partial_+\phi_L+\cos 2\theta\partial_+\phi_R)$$
$$+{\sqrt{k_1/2}
\over 2\pi}\int d^2\sigma (\cos\psi{A_1}_+\partial_-X+ \cos\varphi
{A_1}_-
\partial_+X)$$
$$+{k_1\over 4\pi}\int d^2\sigma (\sin\psi\sin\varphi\sin\alpha\sin\beta
{A_1}_+{A_1}_-+\sin\psi\sin\alpha\cos\beta
{A_1}_+{A_2}_-$$$$+\sin\varphi\sin\beta\cos\alpha{A_2}_+{A_1}_-
+\cos\alpha\cos\beta {A_2}_+{A_2}_-)\cosh 2r$$$$
+(-\sin\psi\sin\varphi\cos\alpha\cos\beta{A_1}_+{A_1}_-
+\sin\psi\cos\alpha\sin\beta {A_1}_+{A_2}_- $$$$
+\sin\varphi\sin\alpha\cos\beta{A_2}_+{A_1}_-
-\sin\alpha\sin\beta
{A_2}_+{A_2}_-)\cos 2\theta $$
$$+{k_1\over 4\pi}\int d^2\sigma ({A_1}_+{A_1}_-(\cos\varphi
  \cos\psi+1) +{A_2}_+{A_2}_-)\eqn\ten$$
It is easy to check that the gauged action preserves the two $U(1)$
local symmetries in \uyfl,\eight. (This is another way to see that the
gauged action is anomaly free.)
Now we pick   two gauge conditions. We take
$$\phi\equiv\phi_R=-\phi_L\;,\;\;\; t\equiv t_R=-t_L
\eqn\nine$$
Among the various (dual)
solutions that are obtained with all the parameters,
our black hole solution is obtained by taking $$\varphi=0\eqn\aghg$$
It must be emphasized that  once $\sin\varphi=0$
we cannot take also $\sin\psi=0$,
since the gauge fixing \nine will not be valid. As we see later,
this is the reason why  this model cannot describe a background without
electromagnetic fields.

Finally,   we are integrating out  the gauge fields $A_1, A_2$
and obtain the following action:
$$ I=\int \D[r,t,\theta,\phi]
 e^{S_{BH}}\det[{-2\pi^2\over k_1\Delta}]\eqn\eleven$$
where   $$\Delta=
  \cos\alpha\cos\beta\cosh 2r-\sin\alpha\sin\beta\cos 2\theta +1
\eqn\twelve$$ and
$$S_{BH}={k_2\over 2\pi}\int d^2\sigma(
\partial_+\theta\partial_-\theta+   {k_1\over k_2}
\partial_+r\partial_-r $$$$-{k_1\over k_2}
 {\sinh^2 r (1+\cos(\alpha+\beta)+2\sin\alpha\sin\beta
\sin^2\theta)\over \Delta}\partial_+t\partial_-t $$$$
+ {\sin^2\theta(2\cos\alpha\cos\beta \cosh^2 r-\cos(\alpha-\beta)
+1)\over \Delta}\partial_+\phi\partial_-\phi
$$$$+2\sqrt {k_1\over k_2}{\sinh^2 r \sin^2 \theta\over \Delta}
(\cos\beta \sin\alpha
\partial_+t\partial_-\phi-  \cos\alpha\sin\beta
\partial_-t\partial_+\phi)$$$$ -{\sqrt{2k_1}
 \over k_2}\tan ({\psi\over 2})
{ \sinh^2 r (2\sin\beta\sin^2\theta
+(\sin\alpha-\sin\beta))\over \Delta}
\partial_+t\partial_-X $$$$
 -\sqrt{2\over k_2} \tan ({\psi\over 2})
 {\sin ^2\theta(
2\cos\beta \cosh^2 r+ (\cos\alpha-\cos\beta))\over \Delta}
\partial_+\phi\partial_-X )$$$$
+{1\over 4\pi}\tan^2 ({\psi\over 2})
\int d^2 \sigma \partial_+X\partial_-X
\eqn\therteen$$The determinant to one loop order is calculated by
following Buscher\Ref\buscher{T.H. Buscher \journal Phys.Lett.
& B201 (88) 466}\Ref\schwartz{A.S. Schwarz and A.A. Tseytlin
\journal Nucl.Phys.&B399 (93) 691} \Ref\kiritsis{E. Kiritsis
\journal Mod.Phys.Lett. &A6 (91) 2871}, and gives rise to the dilaton
term:
 $$\det{-\pi^2\over k_1\Delta}=\exp (-{1\over 8\pi}\int
d^2\sigma\sqrt h R^{(2)}\log \Delta +a)\eqn\fourteen$$
              where $a$ is an arbitrary constant.

Now we  choose $\alpha=\beta$ and denote $$Q=\tan\alpha\eqn\hjhg$$
We further absorb
  $\sqrt{k_1\over k_2}$ in $t$ and absorb
 $\sqrt{2\over k_2} \tan ({\psi\over 2}) $ in $X$.
 (recall that we had to restrict to  $\sin\psi \ne 0$,
 otherwise our gauge fixing is not valid. The rescaling of $X$
 is equivalent
to restricting $\tan({\psi\over 2})=\sqrt{k_2\over 2}$.)
 Finally, we
  redefine the field $r$ as $\hat r=\cosh^2 r$.
Identifying  the sigma model
 \therteen with the string action \bgop we readily see that
the gauged action describes (to one loop order) the following
background (we omitted the hat from $r$ and have an overall
factor of $k_2$):
$$dS^2=-{(r-1)(1+Q^2\sin^2\theta)\over r+Q^2\sin^2\theta}dt^2+
{k_1\over  k_2 r(r-1)}dr^2 + {r\sin^2\theta\over r+Q^2\sin^2\theta}
d\phi^2+d\theta^2 \eqn\fifteen$$
the antisymmetric tensor
 (the "axion field") which has only the $t,\phi$ component
$$B_{t\phi}=2Q{(r-1)\sin^2\theta\over r+Q^2\sin^2\theta}$$
    the electromagnetic vector potential
$$A_t= Q\sqrt{Q^2+1}
{(r-1)\sin^2\theta\over r+Q^2\sin^2\theta}$$
$$A_{\phi}=\sqrt{Q^2+1}
{r\sin^2\theta\over r+Q^2\sin^2\theta}\eqn\sixteen$$
 and the dilaton field is
$$\Phi=\ln(r+Q^2\sin^2\theta)+a\eqn\gep$$
In principle, the remaining part of the metric
 $G_{XX}={1\over 4}$ gives rise to an
additional scalar field (that has only zero mode  in our case),
associated with the following vertex operator
\Ref\vertex{C.G. Callan and Z. Gan, Nucl.Phys.{\bf B272} (1986) 647}
$$\partial X\bar \partial X e^{i(-K_t t+K_r r+K_{\theta}
\theta+K_{\phi} \phi)}\eqn\just$$

This space-time metric \fifteen describes an axisymmetric
 black hole in four dimensions:
 the sphere $r=1$ is the event horizon and $r+Q^2\sin^2\theta=0$ is
 the  singularity,  hidden inside the horizon.
 If one were to interpret $r$ as representing the  radius in polar
coordinates, the fact that there is a singularity at the origin,
$r=0$, only for the angular value $\theta=0,\pi$ appears puzzling.
If we define the metric on the manifold $R^4$ with the origin
removed, we then have incomplete geodesics (such as those on the plane
$\theta=\pi/2$) which terminates at $r=0$ but along which the curvature
remains finite. In fact, this space-time is extendible, and by
defining the coordinate $\tilde r=r+Q^2$ we see that the singularity
has a topology of $S^2\times R$, that is a sphere cross time .
(In the Kerr solution  the singularity
 is at $r^2+a\cos ^2\theta=0$, with $a$ being the total angular
 momentum divided by the mass
\Ref\wald{ R.M. Wald, "General Relativity", The University of
Chicago Press, 1984}.
  In  that solution there is a ring
 singularity (with topology of $S^1\times R$) in the $z=0$ plane.)
 The above singularity is seen by calculating  all
     the  scalar curvatures of the metric (Riemann curvature, Ricci
 curvature, scalar curvature)
 which  all blow up  only at  $r+Q^2\sin^2\theta=0$.
 The expressions for $R_{\mu\nu}$ and the
scalar curvature $R_{\mu}^{\;\mu}$ are given in the appendix.
The classical charged black hole solution,
described by the Nordstrom-Reissner
metric has both inner and outer event horizons. Our dilatonic
charged black hole can thus be regarded as an extremal black hole.

The metric \fifteen is not asymptotically flat. In cartesian coordinates,
at $r\rightarrow \infty  $ the metric approaches (for $k_1=k_2$)
$$dS^2=-(1+Q^2{x^2+y^2\over x^2+y^2+z^2})
dt^2+{dx^2+dy^2+dz^2\over x^2+y^2+z^2}
\eqn\twenty$$
  In other words, it describes  a distribution
of matter all over the space-time.
For $Q=0$ our metric   coincides with that in the
extremal black hole solution in$\lbrack\garfinkle\rbrack$.

 The metric $G_{\mu\nu}$ in \fifteen was read directly from the
 sigma-model action \therteen .  The Einstein metric is obtained by
 conformal rescaling of the sigma-model metric by
  the dilaton field.  In our notation
 $G^E_{\mu\nu }=e^{\Phi}G_{\mu\nu}$, where $\Phi$ is given in
\gep.
Notice that for $Q=0$ and $k_1=k_2$
the Einstein metric after making coordinate transformation
$r\rightarrow r^2$
 is the following:
$$dS^2=-r^2(1-{1\over r^2})dt^2+ (1-{1\over r^2})^{-1}dr^2
+r^2(\sin^2\theta d\phi ^2+d\theta ^2)\eqn\ein$$
and the black hole has  a magnetic field only.

We can calculate all the curvature tensors and scalars in the
Einstein metric and see that there remains a singularity only
at $r+Q^2\sin^2\theta=0$.
(For the formulas of  the transformations
of the curvature tensors and scalar under conformal rescaling
see $\eg$ $\lbrack\wald\rbrack$
.)  Thus the Einstain metric describes a black hole
as well.

{}From the  explicit expression for the electromagnetic
vector field  we obtain the electromagnetic tensor
$F_{\mu\nu}$, defined by $F_{\mu\nu}=\nabla_{\mu}A_{\nu}-\nabla
_{\nu}A_{\mu}.$ Hence,    the
electric ($F_{0,\mu}$) and the magnetic ($F_{i,\mu}$) fields are:
$$E_r= Q\sqrt{Q^2+1}  {(1+Q^2\sin^2\theta)\sin^2\theta\over
 (r+Q^2\sin^2\theta)^2}\eqn\elect$$
$$E_{\theta}= Q\sqrt{Q^2+1} {(r-1)r\sin 2\theta \over
(r+Q^2\sin^2\theta)^2}\eqn\electp$$
 $$B_r= \sqrt{Q^2+1}
 {r^2\sin 2\theta\over (r+Q^2\sin^2\theta)^2}
 \eqn\electo$$
 $$B_{\theta}= \sqrt{Q^2+1}
 {Q^2\sin^4\theta\over (r+Q^2\sin^2\theta)^2}  \eqn\electd$$
If we observe the action \therteen we see that vector potential
is multiplied by $\tan ({\psi\over 2})$, which
   we have absorbed in $X$.\footnote\dagger{We could alternatively
absorb $\sqrt {Q^2+1}$ in $X$ and then the vector potential is
$$A_{t}=Q\tan({\psi\over 2}){(r-1)\sin^2\theta\over r+Q^2\sin^2\theta}
\;\;\;\;\;\;\;\;A_{\phi}=\tan({\psi\over 2}){r\sin^2\theta\over r+Q^2\sin
^2\theta}\;\;\;{\rm and}\;\; G_{XX}=\tan^2({\psi\over 2})/(Q^2+1).$$}
    Only when $\tan ({\psi\over 2})=0$
the black hole has no electromagnetic field. However
 $\sin\psi=0$ invalidates   our gauge fixing \nine (and globally
 reduces  the 5D sigma model action to a 4D one).

Now, the effective action  $\lbrack \callan \rbrack$
of this theory is obtained in the Kaluza-Klein fashion
\Ref\book{T. Appelquist, A. Chodosand P. Freund,
 Modern Kaluza-Klein Theories, Addison-Wesley Publishing
 company, (1987)}as a
dimensional reduction from the five-dimensional effective action.
Denote the scalar field which was introduced as
$G_{XX}=e ^{\varphi}$. Then the effective action can
be written in the following way (we drop the volume element due to
the integration over the fifth dimension and use $\nabla\varphi=0$)
$$S_{eff}=\int d^4x\sqrt{- g} e^{\phi+{1\over 2}\varphi}
(R+(\nabla \phi)^2
-e^{-\varphi}{1\over 4}F^2-{1\over 12}H^2)
\eqn\fifty$$
 where $g$ is the four dimensional Lorentzian metric,
  $F=dA$ and $H=
dB$.
{}From here we see that the  total electric charge is
 $$q_E=\int e^{\phi-{1\over 2}\varphi\;\;}{^*F}d^2 S
 \eqn\fiffour$$
 where the integral is over a 2-sphere at infinity ($^*F$ is
  $F$-dual).
  Thus $$q_E= 4\pi Q\sqrt{Q^2+1}\exp (a) \int_0^{\pi} d\theta\sin^3\theta
 \sqrt{1+Q^2\sin^2\theta}\eqn\rrt$$
The  magnetic charge vanishes  since
 $$q_M=\int Fd^2S=0\eqn\foure $$
Two axionic charges are associated with the action \fifty.
The first one is $$q_{ax}=\int H d^3S=(16\pi/3)Q\eqn\foud$$
The other one ,
 which vanishes in our solution is
$$\tilde q _{ax}=\int F\wedge F=0\eqn\fouc $$(although
locally $F\wedge F\ne 0$).
Thus,  this black hole carries both electric
and axionic charges but has no magnetic charge.
On the other hand, it is well known that the
 equations of motion are  invariant
 under the duality transformation $F\rightarrow  ^*F$
$\lbrack\garfinkle\rbrack$.
 Hence, magnetically charged black hole solution may as well
 be obtained as an equivalent string theory.

 \chapter{The Exact Metric in The Algebraic Hamiltonian Approach}

 In the previous section we have derived a semiclassical background
 by integrating out the gauge fields in the WZW coset model.
The exact to all orders background has $\O({1\over k})$
 corrections, so that in the limit $k_1,k_2\rightarrow \infty$
it reduces to the semiclassical one. Getting the
 precise corrections has a special interest:
The semiclassical metric describes a singularity hidden  by the
 event horizon,
but this is not necessarily a property of
the exact metric. The issue we want to learn is how the space-time
structure of the semiclassical background changes when introducing
all the higher order corrections.
 To find the exact metric that corresponds to the
 solution in section 2 we can use the algebraic Hamiltonian approach
 for cosets $G/H$.
 This method was derived in$\lbrack \verlinde,\bars\rbrack$
 and we first briefly describe it. (For a review see \Ref\itz{I. Bars,
"Curved Spaces Geometry for Strings and Affine Non-Compact Algebras",
preprint USC-93/HEP-B3, to appear in {\it Interface Between
Mathematics and Physics}, Ed. S. T. Yau}.)
We shall concentrate on the closed bosonic strings only.

Consider the Tachyon state $T$, which is
the ground state of the string theory. We denote by $J^G_a$, $J^H_i$ the
currents of the group $G$ and its subgroup $H$, respectively
($a=1,...,\dim G$,
$i=1,...,dim H$) and $J^G_{a,n}, J^H_{i,n}$ are their
``Fourier" components in the Kac-Moody algebra.
 $L_0$ is the zero generator of the Virasoro
algebra. Then the following conditions should be satisfied:
$$(L_0+\bar L_0-2)T=0\;\;;\;\;(J^H_0+\bar J^H_0)T=0\;;\;\;
J^G_nT=\bar J^G_nT=0,\;\;n\ge 1\eqn\sta$$
Here$$L_0={\Delta_G\over k-\tilde c_G}
-{\Delta_H\over k-\tilde c_H}\;\;\;\;\;\;\;\;\;\;\;\;\;\;
\bar L_0={\bar \Delta_G\over k-\tilde c_G}
-{\bar\Delta_H\over k-\tilde c_H}\eqn\ghp$$
where $\Delta_G$,$\Delta_H$ are the Casimir operators in $G$ and in $H$,
\ie $\Delta_G=J^G\cdot J^G$, $\bar\Delta_G={\bar J}^G\cdot{\bar J}^G$,
$\Delta_H=J^H\cdot J^H$, $\bar \Delta_H={\bar J}^H\cdot{\bar J}^H$,
              and $\tilde c_G,\tilde c_H$ are the coexter of $G,H$
respectively.
The second condition in $\sta$ is a remnant of the gauge invariance
$T(h_L g h_R^{-1}) =T(g)$ which demands that the tachyon is a singlet
under the action of the subgroup $H$.

In the algebraic Hamiltonian approach we parameterize the group
elements of $G$ by
$X_{\mu}$, $\mu=1,...,N=\dim G$ and express the currents in terms of
first order differential operators of $X_{\mu}$ which satisfy the
Lie algebra of the group. Then we need to
define gauge invariant coordinates $\tilde X_{\mu}$, $\mu=1,...,D=
\dim
G-\dim H$ and write the Casimir operators in terms of $\tilde X_{\mu}$.
As is well known, the effective action for the Tachyon is
$$S(T)=\int d^DX\sqrt{-G}e^{\Phi}(G^{\mu\nu}\partial_{\mu}T\partial_{\nu}
T-V(T))\eqn\ert$$ where $\Phi$ is the dilaton field and $V(T)$ is the
Tachyon potential. On the other hand, since the Tachyon is completely
defined through the action of the zero modes, its action is equivalent
to $$S(T)=\int d^DX\sqrt{-G}e^{\Phi}(THT-V(T))\eqn\hami$$
 where $H=L_0+\bar L_0$ is the Hamiltonian. Comparing \ert and
 \hami, expressed in terms of $\tilde X_{\mu}$, we obtain
$$L_0+\bar L_0=-e^{-\Phi}{1\over \sqrt{-G}}\partial_{\mu}(e^{\Phi}
\sqrt{-G}G^{\mu\nu}\partial_{\nu})\eqn\qepr$$ from which we find the
exact metric and the exact dilaton field.

Now we return to the gauged action described in section 2.
We  recall that
the background electromagnetic vector
 $A_t, A_{\phi}$ in \sixteen can be written as
$G_{tX}+B_{tX}, G_{\phi X}+B_{\phi X}$ respectively. Thus we are
analyzing a $5\times 5$ metric.
In the previous section, the group elements of $SL(2,R)$ and $SU(2)$
were $h_1=e^{{1\over 2}t_L\sigma_3}e^{r\sigma_1}e^{{1\over 2}t_R
\sigma_3}$ and $h_2
=e^{{i\over 2}\phi_L\sigma_3}e^{i\theta\sigma_1}e^{{i\over 2}\phi_R
\sigma_3}$ respectively, and  the gauge transformation $\uyfl$
amounted to shifting $t_L,t_R,\phi_L,\phi_R$ and $X$ only.
To match theses coordinates
  we  define the following first order differential generators which
satisfy the Lie algebra of $SL(2,R)$ and $SU(2)$.
 Here $J_a$ are the generators of $SL(2,R)$ and $I_a$ are the generators
 of $SU(2)$. ($J_3$ correspond to $-i{\sigma_3\over 2}$ and $I_3$
correspond to ${\sigma_3\over 2}$.)
$$J_3=i\partial_{t_L}\;\;\;\; ;\;\;\;\bar J_3=i\partial_{t_R}$$
$$J_{\pm}=ie^{\pm t_L}({1\over 2}
\partial_r\pm {1\over \sinh 2r}(\partial_{t_R}
-\cosh 2r \partial_{t_L}))$$
$$\bar J_{\pm}=ie^{\pm t_R}({1\over 2}\partial_r\pm
{1\over \sinh 2r}(\partial_{t_L}
-\cosh 2r \partial_{t_R}))\eqn\qsl$$
$$I_3=i\partial_{\phi_L}\;\;\;\; ;\;\;\;\bar I_3=i\partial_{\phi_R}$$
$$I_{\pm}=\pm e^{\mp i\phi_L}({1\over 2}
\partial_\theta\pm{i\over \sin 2\theta}(
\partial_{\phi_R} -\cos 2\theta \partial_{\phi_L}))$$
$$\bar I_{\pm}=\pm e^{\mp i\phi_R}({1\over 2}
\partial_\theta\pm {i\over \sin 2\theta}
(\partial_{\phi_L}
-\cos 2\theta \partial_{\phi_R}))\eqn\sut$$
and we define the generator of the $U(1)$ group by
$$K=\bar K=i\partial_Y$$
In the  coset model  which
described the charged black hole solution \fifteen the $U(1)^2$
gauged subgroup was generated
according to
\uyfl    with $\varphi=0$ and $\alpha=\beta$. So in terms of the
above differential operators
the gauged currents are
 $$\J_1=\sin\psi\sin\alpha J_3+\sqrt{k_1\over
k_2}\sin\psi\cos\alpha
I_3+\sqrt{k_1}\cos\psi K $$
 $$\bar{\J}_1=\sqrt{k_1}\bar K$$
$$\J_2=\cos\alpha J_3-\sqrt{k_1\over k_2}\sin\alpha I_3$$
$$\bar{\J}_2= \cos\alpha\bar J_3+\sqrt{k_1\over k_2}\sin\alpha
\bar I_3\eqn\fkh$$
The central charge of $J_3$ is $k_1$ and the central
charge of $I_3$ is $k_2$, therefore the central charge
of $\J_1$ is $k_1(\sin^2\psi(\sin^2\alpha+\cos^2\alpha)+\cos^2\psi)=k_1$
and the central charge of $\J_2$ is $k_1(\cos^2\alpha+\sin^2\alpha)=k_1$.
Similarly the central charge of $\bar{\J}
_1$,  $\bar{\J}_2$ is also $k_1$.
In the  gauged model we have
$$L_0= {\Delta_{SL(2,R)}\over k_1-2}+{\Delta_{SU(2)}\over
k_2+2}-\partial_Y^2-
{\J_1^2\over k_1}-{\J_2^2\over k_1}\eqn\hkjhg$$
$$\bar L_0= {\bar \Delta_{SL(2,R)}\over k_1-2}+{\bar \Delta_{SU(2)}\over
k_2+2}-\partial_Y^2
-{\bar{\J}_1^2\over k_1}-{\bar{\J}_2^2\over k_1}\eqn\hhg$$
where $$\Delta_{SL(2,R)}=\bar\Delta_{SL(2,R)}=
-{1\over 4}\partial^2_r-{1\over 2}\coth 2r\partial_r$$$$
+{1\over \sinh^2 2r}(\partial^2_{t_L}
-2\cosh2r\partial_{t_L}\partial_{t_R} +\partial^2_{t_R})\eqn\cassl$$
 $$\Delta_{SU(2)}=\bar\Delta_{SU(2)}=
-{1\over 4}\partial^2_{\theta}-{1\over 2}\cot 2\theta\partial_{\theta}
$$$$-{1\over \sin^2 2\theta}(\partial^2_{\phi_L}
-2\cos 2\theta\partial_{\phi_L}\partial_{\phi_R}
+\partial^2_{\phi_R})\eqn\cassu$$
It is easy to see that these Casimir operators produce the
ungauged action  in \four.

Now   we need to find three independent
coordinates $t,\phi,X$ which are linear combinations
 of $t_L,t_R,\phi_L ,\phi_R,Y$ and are gauge invariant, \ie
$$(\J_1+\bar{\J}_1)(t,\phi,Z)=0\;\;\;;\;\;\;(\J_2+\bar{\J}_2)(t,\phi,Z)=0
\eqn\ajsio$$ (namely, it vanishes for each one of the
coordinates separately.
 In the vector gauging we should  replace the + sign by
a - sign.) Thus if the Tachyon is $T(t,\phi,Z)$ it
satisfies the second condition in \sta. (This is like
picking a gauge fixing in the gauged action.)
The exact metric is obtained by substituting $t,\phi,Z$ in
$L_0+\bar L_0$ by using the chain rule. The inverse of the exact
metric is obtained from those terms with quadratic derivatives.
Since $r,\theta$ are unchanged,  $G_{rr}=2(k_1-2)$
and $G_{\theta\theta}=2(k_2+2)$.

Obviously, if  $t,\phi,X$                     fulfil
\ajsio, then any non-vanishing linear combinations of them
are appropriate as well. Different  choices of $(t,\phi,X)$ yield
different (dual) metrics, which are related by similarity
transformations.
We shall return to discuss related subjects in sections 4,5.
In this section, however, we seek the exact metric that correspond
to our solution in section 2. A priory, it is not trivial to guess the
appropriate combinations. Therefore we use the following method. First
we shall calculate the inverse metric of the semiclassical model
in section 2. Since we know that
the exact metric has only $O({1\over k})$ corrections,
we then easily find
the right combinations.
The inverse metric of the {\it semiclassical} model is the following:
(we suppress $k_1, k_2$ factors that were absorbed in $t,\phi$ and
$X$. These factors will come out from the gauge invariant conditions
and we shall absorb them in the coordinates at the end)
$$G^{tt}=-\coth^2 r+\tan^2\alpha\tan^2\theta$$
$$G^{t\phi}=-{1\over \cos^2\alpha}\tan\alpha\tan^2\theta$$
$$G^{tX}={\tan\alpha\over\cos\alpha}\tan^2\theta$$
$$G^{\phi X}=-{1\over \cos^3\alpha}(\tan^2\theta+\cos^2\alpha)$$
$$G^{\phi\phi}= {\tan^2\alpha\over \cosh^2 r}-{1\over \cos^4\alpha}
{-2(1+\cos^2\alpha)\sin^2\alpha\cos 2\theta+(1+\cos^2\alpha)^2+\sin^4
\alpha \over \sin^2 2\theta}$$$$-{\tan^2\alpha\over \cos^2\alpha}$$
$$G^{XX}={1\over \cos^2\alpha}(\tan^2\theta+\cos^2\alpha)
\eqn\afghj$$
{}From these expressions we get the right gauge invariant combinations:
$$t=t_L-t_R-\tan\alpha\sqrt{k_2\over k_1}(\phi_L+\phi_R)\eqn\ttt$$
$$\phi=\tan\alpha(t_L+t_R)-{1\over \cos^2\alpha}\sqrt{k_2\over k_1}
((1+\cos^2\alpha)\phi_R+\sin^2\alpha\phi_L)\eqn\phiphi$$
$$X=\sqrt{k_2\over k_1}{1\over \cos\alpha}(\phi_L+\phi_R)-
{\sin\psi \over \sqrt{k_1}(\cos\psi+1)}Y\eqn\xxx$$
Notice that since $\tan ({\psi\over 2})$
was absorbed in $X$ in \fifteen\sixteen   and $\psi$
disappeared
from the action, we have defined gauge invariant coordinates so that the
exact metric will be independent of $\psi$.
Now we calculate the {\it exact} inverse metric,
Using the chain rule. We obtain the following metric:
$$G^{tt}=-{1\over k_1-2}(\coth^2 r-{2\over k_1})
+{k_2\over (k_2+2)k_1}\tan^2\alpha(\tan^2\theta-{2\over k_2})$$
$$G^{t\phi}=-{k_2\over (k_2+2)k_1}
{1\over\cos^2\alpha}\tan\alpha(\tan^2\theta-{2\over k_2})$$
$$G^{tX}={k_2\over (k_2+2)k_1}
{\tan\alpha\over\cos\alpha}(\tan^2\theta-{2\over k_2})$$
$$G^{\phi X}=-{k_2\over (k_2+2)k_1}
{1\over\cos^3\alpha}(\tan^2\theta+\cos^2\alpha-{2\over k_2}\sin^2
\alpha)$$
$$G^{\phi\phi}= {1\over k_1-2}{\tan^2\alpha\over \cosh^2 r}
-{1\over k_1}{\tan^2\alpha\over \cos^2\alpha}
$$$$-{k_2\over (k_2+2)k_1}{1\over \cos^4\alpha}
{-2(1+\cos^2\alpha)\sin^2\alpha\cos 2\theta+(1+\cos^2\alpha)^2+\sin^4
\alpha \over \sin^2 2\theta} $$
$$G^{XX}={k_2\over (k_2+2)k_1}
{1\over \cos^2\alpha}(\tan^2\theta+\cos^2\alpha-{2\over k_2}\sin^2\alpha)
\eqn\exmet$$
and the dilaton field is
$$\Phi=-{1\over 2}\ln(\Sigma_1\Sigma_2)  \eqn\dill$$   where
$$\Sigma_1=\cosh^2 r+{k_1(k_2+2)\over k_2(k_1-2)}\tan^2
\alpha \sin^2\theta\eqn\sigone$$
$$\Sigma_2=\cosh^2 r+{k_1(k_2+2)\over k_2(k_1-2)}\tan^2
\alpha \sin^2\theta+{2k_1\over (k_1-2)k_2}({k_2\over k_1}-\tan^2\alpha)
\eqn\sigtwo$$
The final step is to calculate the exact metric from its inverse.
Then  we take a pre-factor $2(k_2+2)$ (as we had
in the semiclassical solution),
  absorb $\sqrt{k_1-2\over 2(k_2+2)}$ in $t$, $\sqrt{k_1\over 2k_2}$
in $\phi$ and $\sqrt{k_1-2\over 2(k_2+2)}$ in $X$ and  redefine
$\hat r=\cosh^2 r$. Thus, we obtain the following exact four dimensional
metric (we omitted the hat from $r$):
$$dS^2=-{(r-1)(1+C+Q^2\sin^2\theta)\over r+Q^2\sin^2\theta+C}dt^2
+{(k_1-2)\over (k_2+2) r(r-1)}dr^2$$$$
+{r\sin^2\theta\over r+Q^2\sin^2\theta}d\phi^2
+d\theta^2
\eqn\uyg$$
$$A_t={k_2\over k_2+2}
Q\sqrt{Q^2+{k_1(k_2+2)\over (k_1-2)k_2}}{(r-1)(\sin^2\theta(1+
{2\over k_2})-{2\over k_2})\over r+Q^2\sin^2\theta+C}\eqn\ajhs$$
$$A_{\phi}=\sqrt{Q^2+{k_1(k_2+2)\over (k_1-2)k_2}}
{r\sin^2\theta\over r+Q^2\sin^2\theta}\eqn\dhgsd$$
$$\Phi=-{1\over 2}\ln ((r+Q^2\sin^2\theta)(r+Q^2\sin^2\theta+C))
\eqn\dkjd$$
where $$Q^2={(k_2+2)k_1\over(k_1-2)k_2}\tan^2\alpha\eqn\qqqq$$
and $$C={2\over k_1-2}(1-{k_1\over k_2}\tan^2\alpha)\eqn\cccc$$
(Notice that when $k_1<2$ we should take $\vert k_1-2\vert$ in
$C,Q$ since we have absorbed $\sqrt{k_1-2}$ in t,X. Thus
 $Q^2\ge 0$.)
It is easy to see that for $k_1,k_2\rightarrow\infty$ the exact
solution is precisely the semiclassical one, however,
 for finite $k_1$,$k_2$ the space-time might
change drastically.
In the appendix we have given the expressions for the
Ricci tensor and the scalar curvature of this metric.
It can be seen that $r=1$ is the event horizon, as in the semiclassical
solution.
The metric is singular in three cases:
(i) when $\Sigma_1= r+Q^2\sin^2\theta=0$.\nextline
 (ii) when $\Sigma_2=r+Q^2\sin^2\theta+C=0$.\nextline
(iii) when $1+C+Q^2\sin^2\theta=0$.\nextline
 (Notice that $C+Q^2={1\over k_1-2}(k_1\tan^2\alpha+2)$)\nextline
(a) For $C>-1$ ($\tan^2\alpha<{k_2\over 2}$):
 The singularity is hidden by the event horizon.
In this case the solution describes a black hole.\nextline
(b) For $C=-1$: The black hole singularity extends up to the horizon.
In addition, there exist
a naked {\it string} singularity (at $\sin\theta=0$) that crosses the
event horizon.\nextline
(c) For  $C<-1$: The black hole singularity extends outside the
event horizon and becomes a naked singularity. In addition, there
is a singularity on two
 cone surfaces
 ($\theta=arc\sin({\sqrt{\vert 1+c\vert}\over Q})$ and
 $\theta=\pi-arc\sin({\sqrt{\vert 1+c\vert}\over Q})$)
which cross the event horizon and become naked.
\nextline
Hence, we reach the following conclusion: Unlike in the 2D black hole
case, the
exact metric is singular  (for any choice of the gauge parameter
$\alpha$), and furthermore,  for a certain
range of the gauge parameter (where
$\tan^2\alpha\ge{k_2\over 2}$)    the semiclassical action
describes a black hole while the exact one describes a non-physical
space-time.

Finally,   when $\tan^2\alpha={k_2\over k_1}$ the
semiclassical metric and dilaton and the
exact metric and dilaton
 are identical, up to shifting $k_1\rightarrow
k_1-2$ and $k_2\rightarrow k_2+2$.   It is
not possible to derive the antisymmetric tensor by the algebraic
Hamiltonian approach. However, in the sigma model where we have
integrated out the gauge fields, the antisymmetric tensor
 has components $B_{X\phi}=G_{X\phi}$ , $B_{X t}=G_{Xt}$ and
$\sqrt{Q^2+1\over k_1}B_{t\phi}=G_{zt}$. We conjecture that because
of the construction of the sigma model, the first two equalities
remain also in the exact solution and the last one is corrected by
a $C$  dependence   (like $A_t$),  so that when $C=0$
only $A_t,B_{t\phi},B_{tz}$ have $\O({1\over k})$ corrections.

\chapter{Exact  Dual Models}

In section 3 we have used the algebraic Hamiltonian approach to calculate
the exact metric and dilaton field
that correspond to our black hole solution in section 2.
This means the following: We have used  specific generators
for the $U(1)^2$  gauged group (specific gauging)
 that matched the gauging in the WZW sigma model
and for these generators  we have used
specific gauge invariant coordinates that matched the classical
solution.
 In order to get {\it all} the dual metrics we should
consider all the different anomaly free gaugings and all different
gauge invariant combinations for each gauging.
It is easy to see that
different gauge invariant coordinates for one particular gauging
 correspond to a constant coordinate transformation.
The aim of this section is to derive a formula for all the dual
metrics. We will see that all the dual models are related by
$O(3,3)$ symmetry transformations,
of which the semiclassical limit is well known.

In the model we were using in section 3 we gauged a $U(1)^2$
subgroup whose generators correspond to the transformations\uyfl.
Instead of looking for all other anomaly free generators we shall use the
following method:
Consider the $L_0+\bar L_0$ operator we have used for the gauged
model in \hkjhg\hhg\fkh.  This can be written as
$$L_0+\bar L_0=2{\Delta_{SL(2,R)}\over k_1-2}+2{\Delta_{SU(2)}
\over k_2+2}-2\partial^2_Y-{1\over k_1}\sum_{i=1,2}(\J_i^2+
\bar {\J}_i^2)$$$$=
(\partial_{t_L},\partial_{\phi_L},\partial_Y)
(G^{LL}-\J^{LL})\left (\matrix{\partial_{t_L}\cr
\partial_{\phi_L}\cr\partial_Y\cr}
\right )+(\partial_{t_L},\partial_{\phi_L},\partial_Y)
G^{LR}\left (\matrix{\partial_{t_R}\cr\partial_{\phi_R}\cr}\right )
$$$$+ (\partial_{t_R},\partial_{\phi_R})(G^{RR}-\J^{RR})
\left (\matrix{\partial_{t_R}\cr\partial_{\phi_R}\cr}\right)-
{\partial_r^2+2\coth 2r\partial_r\over 2(k_1-2)}
-{\partial_{\theta}^2+2\cot 2\theta\partial_{\theta}\over 2(k_2+2)}
\eqn\ghj$$
where $G^{LL},G^{LR},G^{RR}$ are obtained from the Casimir operators
of the group $SL(2,R)\times SU(2)\times U(1)$
and $\J^{LL},\J^{RR}$ are obtained from ${1\over k_1}
\sum(\J_i^2+\bar{\J}_i^2)$.
(Here $G^{LR}$ is a $(3\times 2)$ matrix with zeros in the last line.)
Denote $$\left (\matrix{t\cr\phi\cr X\cr}\right)=
A\left (\matrix{t_L\cr\phi_L\cr Y\cr}\right)+\tilde B
\left (\matrix{t_R\cr\phi_R\cr}\right)\eqn\jpsc$$
where $A,\tilde B$ are two $3\times 3$ and
$3\times 2$ matrices, respectively, obtained from \ttt -\xxx.
 The inverse metric is of course
block diagonal and in the block of $t,\phi,X$ it is
$$G^{-1}=-A^T G^{LL}A-A^TG^{LR}\tilde B-\tilde B^TG^{RR}\tilde B
+A^T\J^{LL}A+\tilde B^T\J^{RR}\tilde B\eqn\kjhjah$$
For any constant matrices $O_1$  and $\tilde O_2$ which are $O(3)$
and $O(2)$ matrices, respectively, the transformation
$$\left (\matrix{t_L\cr \phi_L\cr X\cr}\right )\rightarrow O_1
\left (\matrix{t_L\cr \phi_L\cr X\cr}\right )\;\;\;\;\;;\;\;\;\;
\left (\matrix{t_R\cr\phi_R\cr}\right )\rightarrow \tilde O_2
\left (\matrix{t_R\cr\phi_R\cr}\right )\eqn\lhpp$$
$$G^{LL}\rightarrow O_1G^{LL}O_1^T\;\;\;\;;\;\;
G^{LR}\rightarrow O_1G^{LR}\tilde O_2^T\;\;\;\;;\;\;
G^{RR}\rightarrow \tilde O_2G^{RR}
\tilde O_2^T\eqn\ajklg$$  leaves the Casimir
operators of the ungauged model
unchanged. We can now express the ungauged model with the
rotated  coordinates $t'_L,t'_R,\phi'_L,\phi'_R,X'$ (rotate the Casimir
operator of the ungauged group)
 while gauging the subgroup generated
 with the same generators (not rotate the generators), $\eg$
$\J_1=i(\sin\psi\sin\alpha\partial_{t'_L}+\sqrt{k_1\over k_2}\sin\psi
\cos\alpha\partial_{\phi'_L}+\sqrt{k_1}\partial_{Y'})$, etc'.
 This is of
course still an anomaly free gauging. Thus $A,B$ and $\J^{LL},\J^{RR}$
are unchanged and we get dual models with
$$G^{-1}=-A^T O_1G^{LL}O_1^TA-A^TO_1G^{LR}
\tilde O_2^T\tilde B-\tilde B^T\tilde O_2G^{RR}\tilde O_2^T\tilde B
+A^T\J^{LL}A+\tilde B^T\J^{RR}\tilde B\eqn\kh$$
 However, with this method we can find only dual models
that are related by $O(3)\times O(2)$ duality, while we expect our
model to possess $O(3)\times O(3)$ symmetry (since the background is
independent of the 3 coordinates $t,\phi,X$).
 In order to see the full symmetry we need to
use an equivalent model. We consider the model $SL(2,R)\times SU(2)
\times U(1)^2/U(1)^3$. The two $U(1)$ groups are defined by $X_L, X_R$
so that in the notations of section 3 we have the currents
$$K=i\partial_{X_L}\;\;\;;\;\;\;\bar K=i\partial_{X_R}$$
and  $\partial^2_Y$ is replaced by $\partial^2_{X^L}+\partial^2_{X^R}$
in the operator $L_0+\bar L_0$ of the ungauged model
used in section 3.
We need to define three left and right generators that fulfil the
anomaly free condition \six.
We shall choose the generators that produce exactly the same metric
we derived in the $SL(2,R)\times SU(2)\times U(1)/U(1)^2$ model
and then show how to derive all other dual models.
We write the group elements as $g=diag(h_1,h_2,e^{iX_L},e^{iX_R})$,
where as in section 2 $h_1\in SL(2,R),\;h_2\in SU(2)$.
Now we gauge the $U(1)^3$ currents that correspond to the following
generators:
$$T_{1,L}= {1\over \sqrt {2}}diag (\sin\alpha{\sigma_3\over 2},
  \sqrt{k_1\over k_2}\cos\alpha{i\sigma_3\over 2},
i\sqrt{k_1\over 2},0)$$
$$T_{1,R}= diag ( 0,0,0,
i\sqrt{k_1\over 2})$$
$$T_{2,L}= diag ( \cos\alpha{\sigma_3\over 2},
-\sqrt{k_1\over k_2}\sin\alpha{i\sigma_3\over 2},0,0)$$$$
T_{2,R}= diag (\cos\alpha{\sigma_3\over 2},
\sqrt{k_1\over k_2}\sin\alpha{i\sigma_3\over 2},0,0)$$
$$T_{3,L}=  diag (0,0,i\sqrt{k_1\over 2} ,0)$$$$
  T_{3,R}= {1\over \sqrt{2}} diag (\sin\alpha{\sigma_3\over 2}
,-\sqrt{k_1\over k_2}\cos\alpha{i\sigma_3\over 2},0,
i\sqrt{k_1\over 2})\eqn\syve$$
In the axial gauging this  corresponds to the three constraints:
$$0=\J_1+\bar{\J}_1={1\over\sqrt{2}}(\sin\alpha J_3+\sqrt{k_1\over k_2}
\cos\alpha I_3+\sqrt{k_1}K)+\sqrt{k_1}\bar K$$
$$0=\J_2+\bar{\J}_2=\cos\alpha J_3-\sqrt{k_1\over k_2}
\sin\alpha I_3+\cos\alpha\bar J_3+\sqrt{k_1\over k_2}
\sin\alpha\bar I_3$$
$$0=\J_3+\bar{\J}_3=\sqrt{k_1}K+
{1\over\sqrt{2}}(\sin\alpha \bar J_3-\sqrt{k_1\over k_2}\cos\alpha
\bar I_3+\sqrt{k_1}\bar K)\eqn\kjhg$$
We take the gauge invariant coordinates that match those
we used in\xxx, so that we only replace the dependence on $Y$ by
a dependence on $X^L,X^R$.
$$t=t_L-t_R-\tan\alpha\sqrt{k_2\over k_1}(\phi_L+\phi_R)\eqn\ttt$$
$$\phi=\tan\alpha(t_L+t_R)-{1\over \cos^2\alpha}\sqrt{k_2\over k_1}
((1+\cos^2\alpha)\phi_R+\sin^2\alpha\phi_L)$$$$
-{4\over \sqrt{2k_1}\cos
\alpha}(X_L-{1\over \sqrt{2}}X_R)\eqn\hjgfsdc$$
$$X=\sqrt{k_2\over k_1}{1\over \cos\alpha}(\phi_L+\phi_R)+
{1\over \sqrt{k_1}(\sqrt{2}-1)}(X_L-X_R)\eqn\xpx$$
which we shall write as $$\left (\matrix{t\cr\phi\cr X\cr}\right)=
A\left (\matrix{t_L\cr\phi_L\cr X_L\cr}\right)+B
\left (\matrix{t_R\cr\phi_R\cr X_R\cr}\right)\eqn\jgsc$$
and   $A,B$ are the two corresponding $3\times 3$ matrices.
It is easy to see that this model yields precisely the metric we
found in section 3: Using the constraints  $\J_i^2=\bar {\J}_i^2$
it can be seen that when substituting $t,\phi,X$
all the contributions from $\partial_{X^L}$
and $\partial_{X^R}$ cancel out, as in the case in section 3
where all derivatives $\partial_Y$  cancelled out.

At this stage we can readily derive all the exact
dual models. Here we shall write in details.
 First we write $L_0+\bar L_0$ in the following way:
$$L_0+\bar L_0=(\partial_{t_L},\partial_{\phi_L},\partial_{X_L})(G^{LL}
-\J^{LL})
\left (\matrix{\partial_{t_L}\cr \partial_{\phi_L}\cr\partial_{ X_L}\cr}
\right ) +(\partial_{t_L},\partial_{\phi_L},\partial_{X_L})
G^{LR}\left (\matrix{\partial_{t_R}\cr\partial_{\phi_R}\cr
\partial_{X_R}\cr}\right )$$$$+
(\partial_{t_R},\partial_{\phi_R},\partial_{X_R})
(G^{RR}-\J^{RR})\left (\matrix{\partial_{t_R}\cr\partial_{\phi_R}\cr
\partial_{X_R}\cr  }\right)
-{\partial_r^2+2\coth 2r\partial_r\over 2(k_1-2)}
-{\partial_{\theta}^2+2\cot 2\theta\partial_{\theta}\over 2(k_2+2)}
\eqn\nrj$$
where $G^{LL}$,$G^{LR}$,$G^{RR}$ correspond to the Casimir operator
of the ungauged model and $\J^{LL}$,$\J^{RR}$ correspond to the gauged
$U(1)^3$ currents.
$$G^{RR}=G^{LL}=\left (\matrix {{2\over (k_1-2)\sinh^2(2r)}& &\cr
                      &{-2\over (k_2+2)\sin^2(2\theta)}& \cr
                        & &-2\cr}\right) $$$$
G^{LR}=\left(\matrix{{-4\cosh(2r)\over (k_1-2)\sinh^2(2r)}& &\cr
              &{4\cos(2\theta)\over (k_2+2)\sin^2(2\theta)}&\cr
              & &0\cr}\right)\eqn\matfhg$$
 $$\J^{LL}=-{1\over 2k_1}\left (\matrix{\cos^2\alpha+1&-\sqrt{k_1\over
k_2}\sin\alpha\cos\alpha&\sqrt{k_1}\sin\alpha\cr
-\sqrt{k_1\over k_2}\sin\alpha\cos\alpha&{k_1\over k_2}(\sin^2\alpha+1)&
{k_1\over \sqrt{k_2}}\cos\alpha\cr
\sqrt{k_1}\sin\alpha&{k_1\over \sqrt{k_2}}\cos\alpha& 3k_1\cr}\right )
\eqn\khhhj$$
 $$\J^{RR}=-{1\over 2k_1}\left (\matrix{\cos^2\alpha+1&\sqrt{k_1\over
k_2}\sin\alpha\cos\alpha&\sqrt{k_1}\sin\alpha\cr
\sqrt{k_1\over k_2}\sin\alpha\cos\alpha&{k_1\over k_2}(\sin^2\alpha+1)&
-{k_1\over \sqrt{k_2}}\cos\alpha\cr
\sqrt{k_1}\sin\alpha&-{k_1\over \sqrt{k_2}}\cos\alpha& 3k_1\cr}\right )
\eqn\hgjkds$$
Now the Casimir operator of the ungauged group is invariant under
the $O(3)\times O(3)$ transformation
$$\left (\matrix{t_L\cr \phi_L\cr X_L\cr}\right )\rightarrow O_1
\left (\matrix{t_L\cr \phi_L\cr X_L\cr}\right )\;\;\;\;\;;\;\;\;\;
\left (\matrix{t_R\cr\phi_R\cr X_R\cr}\right )\rightarrow O_2
\left (\matrix{t_R\cr\phi_R\cr X_R\cr}\right )$$
$$G^{LL}\rightarrow O_1G^{LL}O_1^T\;\;\;\;;\;\;
G^{LR}\rightarrow O_1G^{LR}O_2^T\;\;\;\;;\;\;
G^{RR}\rightarrow O_2G^{RR}O_2^T\eqn\ajkhg$$  where
$O_1$ and $O_2$ are two constant
$O(3)$ matrices. From now we just repeat the steps
from \kjhjah to \kh, \ie rotate $G^{LL},G^{LR},G^{RR}$ while gauging
the anomaly free subgroup generated by\syve.
Therefore we get an expression for the $t,\phi,X$ components of the
metric in all the dual models:
$$G^{-1}=-A^T O_1G^{LL}O_1^TA-A^TO_1G^{LR}O_2^TB-B^TO_2G^{RR}O_2^TB
+2A^T\J^{LL}A\eqn\kjah$$
where we  used  $A^T\J^{LL}A=B^T\J^{RR}B$.
The other generator of the $O(3,3)$ symmetry are: coordinate
transformations  $(t,\phi,X)\rightarrow (t,\phi,X)C^T$where $C$
is a constant $GL(3,R)$ matrix- this amounts to
 transforming $G\rightarrow C^T GC$,
and a constant shift of the antisymmetric tensor.
(Notice that by   a similarity transformation
one can diagonalize $A^T\J^{LL}A$ and $B^T\J^{RR}B$ to become
$-{\tilde c_G\over k_1}\oe$, where $\oe$ is the unit matrix.)
Thus we extended the $O(d,d)$ symmetry to the exact case.

In particular, we can obtain the axial-vector duality. This duality
was investigated in the sigma model of $U(1)^d$ gauged WZW
in$\lbrack\kiritsis,\amit\rbrack$.
As mentioned before,
in the algebraic Hamiltonian approach the gauge invariance conditions
for abelian gauging
are $$\J_i\pm \bar{\J}_i=0$$ where the $+$ sign correspond to the
axial gauging and the  $-$ sign to the vector gauging.
In particular one can interchange axial-vector gauging by taking
$O_1=-O_2=\oe$ in \kjah, where $\oe$ is the unit matrix.

In the rest of this section we shall examine the vector gauging
that correspond to the generators we used  in section 3 for
the axial gauging
(\ie use the same currents $\J_1,\J_2,\bar {\J_1},\bar{\J}_2$ in
\fkh ).   These two CFT's are completely equivalent$\lbrack
\amit\rbrack$.
As is easily seen, the ${1\over k_1}$ term in $L_0+\bar L_0$
is the same as in  the axial gauging.
The     exact inverse metric in the vector gauging is
$$G^{tt}=-{1\over k_1-2}(\tanh^2 r-{2\over k_1})
+{k_2\over (k_2+2)k_1}\tan^2\alpha(\cot^2\theta-{2\over k_2})$$
$$G^{t\phi}=-{k_2\over (k_2+2)k_1}
{1\over\cos^2\alpha}\tan\alpha(\cot^2\theta-{2\over k_2})$$
$$G^{tX}={k_2\over (k_2+2)k_1}
{\tan\alpha\over\cos\alpha}(\cot^2\theta-{2\over k_2})$$
$$G^{\phi X}=-{k_2\over (k_2+2)k_1}
{1\over\cos^3\alpha}(\cot^2\theta+\cos^2\alpha-{2\over k_2}\sin^2
\alpha)$$
$$G^{\phi\phi}= {1\over k_1-2}{\tan^2\alpha\over \sinh^2 r}
-{1\over k_1}{\tan^2\alpha\over \cos^2\alpha}
$$$$-{k_2\over (k_2+2)k_1}{1\over \cos^4\alpha}
{2(1+\cos^2\alpha)\sin^2\alpha\cos 2\theta+(1+\cos^2\alpha)^2+\sin^4
\alpha \over \sin^2 2\theta} $$
$$G^{XX}={k_2\over (k_2+2)k_1}
{1\over \cos^2\alpha}(\cot^2\theta+\cos^2\alpha-{2\over k_2}\sin^2\alpha)
\eqn\exme$$
and the dilaton field is
$$\Phi=-{1\over 2}\ln(\Sigma_1\Sigma_2)  \eqn\dll$$   where
$$\Sigma_1=\sinh^2 r+{k_1(k_2+2)\over k_2(k_1-2)}\tan^2
\alpha \cos^2\theta\eqn\sione$$
$$\Sigma_2=\sinh^2 r+{k_1(k_2+2)\over k_2(k_1-2)}\tan^2
\alpha \cos^2\theta+{2k_1\over (k_1-2)k_2}({k_2\over k_1}-\tan^2\alpha)
\eqn\sitwo$$
We see that the only difference between this solution and
 the axially gauged
solution in section 3 \exmet\dill
is a replacement  $\cos\theta\leftrightarrow \sin\theta$, $\cosh r
\leftrightarrow \sinh r$. Therefore, the models that correspond to
the axial and the vector gaugings   are self dual: One can transform
from each other by a shift $\theta\rightarrow \theta+{\pi\over 2}
\;\;\;r\rightarrow r+i{\pi\over 2}$.

 Finally, we shift $\theta$ to  $\theta+\pi/2$ and redefine $\sinh^2 r
\rightarrow r$. Then the metric with a pre-factor $2(k_2+2)$ (and
with the appropriate absorption of constants in $t,\phi,X$) is
$$dS^2=-{(r+1)(1+C+Q^2\sin^2\theta)\over r+Q^2\sin^2\theta+C}
dt^2 +{k_1-2\over k_2+2}{dr^2\over r(r+1)}$$
$$+{r\sin^2\theta\over r+Q^2\sin^2\theta}d\phi^2
+d\theta^2 \eqn\kjhgf$$
where $C,Q$ are defined in\cccc\qqqq, and the electromagnetic vector has
the $t,\phi$ components.
For $C\ge 0$
this metric describes a naked singularity at $r+Q^2\sin^2\theta=0$.
(When $C=0$ the exact metric is the same as the semiclassical one.)
For $0>C>-1$ there is a naked singularity at $r+Q^2\sin^2\theta+C=0$.
For $C=-1$ there exist additional naked
string singularity (at $\sin\theta=0$).
This singularity does not exist  in the semiclassical limit.
For $ C< -1$
there is, in addition,  a new naked
singularity at the two cone surfaces
$\theta_1=arc\sin (\sqrt{-1-C}/Q)$ and $\theta_2=\pi-\theta_1$.

\chapter{Exact $O(d,d)$ transformations of the metric and the dilaton}

The $O(d,d)$ symmetry (duality)
appears when the background in independent
of $d$ of the $D$ space-time coordinates. It can be seen at the
classical level that there is a symmetry transformation that can
be applied on the background $G_{\mu\nu}$, $B_{\mu\nu}$, accompanied
by a transformation of the dilaton field, that leaves the one loop
effective action unchanged$\lbrack \veneziano,\gasperini\rbrack$.
The symmetry transformation can be derived also by gauging a $U(1)^d$
subgroup in a sigma-model with $D+2d$ target space dimensions
with $2d$ Killing vectors$\lbrack
\giveon\rbrack$ (\ie the ungauged
background is independent of the $2d$ coordinates from which we gauge
out $d$) and also by means
of string field theory$\lbrack\sen,\hassan\rbrack$.
The latter two approaches
 gives the one loop order duality transformations based on conformal
field theories. Here we shall interpret the sigma-model approach
in $\lbrack\giveon\rbrack$
to the exact action by the algebraic Hamiltonian approach.

Consider a $D$ dimensional background that  is independent
of $d$ coordinates which we denote by $Y_i$, $i=1,...,d$ and
the rest of the coordinates are denoted by $X_{\mu}$, $\mu=1,...,D-d$.
In this section we shall consider the $O(d,d)$ transformations
for the case when the target space metric satisfy $G_{i\mu}=0$.
The generalization to the case $G_{i\mu}(X)\ne 0$ can be established
as well\Ref\dudi{D. Gershon, "Exact $O(d,d)$ Transformations in WZW
Models", preprint TAUP-2129-93}. We shall denote
$G_{ij}$ by $G$ and $G_{\mu\nu}$ by $\tilde G$.

We shall consider a group $G$ WZW model with level $k$  that is described
by the following sigma model:
$$S={k\over 8\pi}\int d^2\sigma (\tilde G_{\mu\nu}(X)
\partial_+X^{\mu}\partial_-X^{\nu}+\partial_+\theta_1^i\partial_-
\theta_1^i +\partial_+\theta_2^i\partial_-\theta_2^i +2E_{ij}(X)\partial
_+\theta_1^i\partial_-\theta_2^j\eqn\ghgh$$
The action is described by a target space with $D+d$ dimensions
with $X_{\mu}$, $\mu=1,..,D-d$, and $\theta_1^i,\theta_2^i$
$i=1,...d$. Now we want to gauge
the $U(1)^d_L\times U(1)^d_R$ subgroup, that correspond to the
holomorphic conserved currents
$$J^i=\partial_+\theta_2^i+E_{ji}\partial_+\theta_1^j$$
$$\bar J ^i=\partial_-\theta_1^i+E_{ij}\partial_-\theta_2^j\eqn\ahgk$$
Let us represent this model by the algebraic Hamiltonian approach.
The ungauged WZW is exact (up to a shift $k\rightarrow k-\tilde c_G$). Reading
the casimir operator from the ungauged model \ghgh it can be written as
$$-\Delta=-\bar \Delta=
 K^{\mu}(X)\partial_{X^{\mu}}+F^{\mu\nu}(X)\partial_
{X^{\mu}}\partial_{X^{\nu}}+(\oe -EE^T)^{-1}_{ij}\partial_{\theta_1^i}
\partial_{\theta_1^j}$$$$ +(\oe-E^TE)^{-1}_{ij}
\partial_{\theta_2^i}\partial_{\theta_2^j})
-2(E(\oe-E^TE)^{-1})_{ij}\partial_{\theta_1^i}\partial_{\theta_2^j}
\eqn\uhuh$$where  $\oe$ is the $(d\times d)$ unit matrix.
We parameterize the $U(1)$ gauged currents by the commuting generators
$$J_i=i\partial_{\theta_1^i}\;\;\;;\;\;\;\bar J_i
=i\partial_{\theta_2^i}\eqn\skjh$$ and take the gauged currents to be
$\J_i=J_i$ and $\bar{\J}_i=\bar J_i$.
Obviously they correspond to an
anomaly free gauging.
The coset model $G/U(1)^d$ correspond to
$$L_0+\bar L_0={2\Delta\over k-\tilde c_G}-{1\over k}\sum_{i=1}^d(\J_i^2
+\bar{ \J}_i^2))\eqn\jhgf$$
We shall use the axial gauging. Define the gauge invariant
coordinates $$Y^i=\theta_1^i+\theta_2^i\eqn\dvid$$
Substituting $Y^i$ in \jhgf
 we obtain the following $(D\times D)$ metric of the  coset model:
$$G_{G/H}^{-1}=\left (\matrix {\tilde G^{-1}&0\cr
                       0&G^{-1}\cr}\right )$$
 where
$$G^{-1}={2\over k-\tilde c_G}\lbrack
(E^{-1}-E^T)^{-1}(\oe+{1\over 2}(E^{-1}+E^T))
+({E^T}^{-1}-E)^{-1}(\oe+{1\over 2}({E^T}^{-1}+E))$$$$
+{\tilde c_G\over k}\oe\rbrack\eqn
\gfghj$$and $\tilde G$ is unchanged (\ie obtained from
$F^{-1}$).
Now,  the Casimir operator of the (ungauged) group $G\uhuh$
(alternatively, the ungauged  action \ghgh) is invariant under the
transformations $$\theta_1\rightarrow O_1\theta_1\;\;\;;
\;\;\;\theta_2\rightarrow O_2\theta_2
\;\;\;\;\;;\;\;\;E\rightarrow O_1 EO_2^T
\;\;\;(E^T\rightarrow O_2 E^TO_1^T)\eqn\jhga$$
where $O_1, O_2$ are two constant $O(d)$ matrices.
This transformation is just a duality transformation, however,
if the coordinates $\theta_1^i,\theta_2^i$ are compactified, one
should restrict to $O(d,\Z)$ matrices in order to preserve the
periodicity. In the latter case, if we take
 general $O(d,\R)$ matrices the action is still
 conformal but not equivalent to the original one.
As we did in section 4, we rotate the
coordinates $\theta_1,\theta_2$ independently, then
express the Casimir operator of $G$
in terms of the rotated coordinates,
but unchange the generators of  the gauged subgroup (namely,
$\J_i=i\partial_{\theta'^i_1}$ and $\bar{\J}_i=
i\partial_{\theta'^i_2}$). Thus the only change in $G^{-1}$ is
$E\rightarrow O_1 E O_2^T$ ($E^T\rightarrow O_2E^TO_1^T$).
Hence, we obtain the exact transformation of the metric through
the transformation of $E$. Of course this $O(d)\times O(d)$
duality transformation
is accompanied by a transformation of the exact antisymmetric tensor
which we  know only to one loop order.
The transformation of the
dilaton term can be found from the first order differential
operators which are not changed under the $O(d)\times O(d)$
duality.
Therefore it is easy to see that $e^{\Phi}\sqrt{G}$ must be
invariant under the duality transformation, \ie
$$\Phi'=\Phi +{1\over 2}\ln({\det G\over \det G'})\eqn\htfga$$
where $G'$ is the transformed metric. This
 is the same transformation as in the semiclassical limit.
The fact that $e^{\Phi}\sqrt{G}$ is independent of $k$ was pointed
out in $\lbrack\bars\rbrack$.
The matrix $E$ is general. Starting with an exact metric that correspond
to some matrix $E$ implies that there is a larger conformal theory from
which the coset model can be obtained (since $S_{G/H}=S_G-S_H$).
Thus the procedure is general for all
 models where the metric  $G_{\mu\nu}$
 is independent of $Y^{i}$ and $G_{i\mu}=0$.

Obviously,
one could choose other invariant coordinates $\tilde Y_i=
C_{ij}Y_i$ in \dvid, where $C$ is a $GL(d,R)$ matrix. Then
$G^{-1}\rightarrow C^TG^{-1}C$
so the higher orders corrections
to the inverse metric is not necessarily diagonal
(like the metric we derived in section 3).
But  given an exact metric, one can
diagonalize the ${\tilde c_G\over k}$ correction by a constant coordinate
transformation. The important point is that this
term- the ${1\over k}$ correction of the inverse metric
with respect to the one loop order-
is unchanged under the $O(d)\times O(d)$ duality.

We return now to the notation  we were using in section 4, but write
the $k$ dependence explicitly.
In general,  when gauging a $U(1)^d$ subgroup in an action with
$2d$ isometries, with   $G_{i\mu}=0$, taking the $d$ gauge invariant
coordinates as the vector
$Y=A\theta^L+B\theta^R$ (where $A,B$ are $(d\times d)$  matrices),
the exact inverse metric for the $Y_i$ components is
$$G^{-1}=-{1\over k-\tilde c_G}( A^TG^{LL}A+A^TG^{LR}B+B^TG^{RR}B)
+{1\over k}(A^T\J^{LL}A+B^T\J^{RR}B)$$$$=
-{1\over k-\tilde c_G}\lbrack A^T(G^{LL}-\J^{LL})A+A^TG^{LR}B+
B^T(G^{RR}-\J^{RR})B$$$$
+{\tilde c_G\over k}(A^T\J^{LL}A+B^T\J^{RR}B)\rbrack=
{1\over k-\tilde c_G}(G^{-1}_{classical}
-{2\tilde c_G\over k} A^T J^{LL}A)\eqn\kjhgs$$
where $G^{LL},G^{LR},G^{RR}$ correspond to the Casimir operator
of the ungauged group and $\J^{LL},\J^{RR}$ correspond to the gauged
currents.
We used $A^T\J^{LL}A=B^T\J^{RR}B$ and took the classical
metric with a pre-factor $k$. (The classical metric is obtained
by plugging $\tilde c_G=\tilde c_H=0$.)
Under the $O(d)\times O(d)$ duality only $G^{LL},G^{LR},G^{RR}$ change.
Thus only the semiclassical part of the inverse metric changes. Moreover,
all the  semiclassical backgrounds which are obtained by $
O(d)\times O(d)$ transformations
can be obtained also by  different gaugings of  the ungauged
action$\lbrack\giveon\rbrack$ (\ie picking different generators for the
gauged subgroup).
So one can apply the one loop transformation on the classical part
of the inverse metric while leaving the ${1\over k}$ correction
unchanged and get  exact $O(d)\times O(d)$ dual models.
Writing $$G^{-1}_{exact}={1\over k-\tilde c_G}
(G^{-1}_{classical}+2{\tilde c_G\over k}
C^TC)\eqn\sjh$$ under the exact $O(d)\times O(d)$ duality the
transformation is $${G'}^{-1}_{exact}=
{1\over k-\tilde c_G}({G'}_{classical}^{-1}
+2{\tilde c_G\over k}C^TC)\eqn\kujas$$
$$\Phi'_{exact}=\Phi_{exact}+{1\over 2}\ln ({\det G_{exact}\over
              \det {G'}_{exact}})\eqn\lkh$$
where ${G'}_{classical}^{-1}$ is the dual classical metric. We see
that in order to find exact dual models we do not need to have the
exact antisymmetric tensor!  Denote  the one loop order limit of $G,B$
by $\hat G, \hat B$ then the general $O(d)\times O(d)$ transformation
to one loop order is$\lbrack\sen\rbrack$
$${G'}_{classical}^{-1}={1\over 4}((O_1+O_2)\hat G^{-1}(O_1+O_2)^T+
(O_1-O_2)(\hat G-\hat B\hat G^{-1}\hat B)(O_1-O_2)^T$$$$
-(O_1+O_2)\hat G^{-1}\hat B(O_1-O_2)^T
+(O_1-O_2)\hat B\hat G^{-1}(O_1+O_2))\eqn\hgjdh$$
which we now substitute in \kujas.

 The rest of the $O(d,d)$ generators apply as in the semiclassical
 limit. These are
 $G_{exact}\rightarrow C^TG_{exact}C$, where $C$ is a $GL(d,R)$
constant matrix, and  constant shifts of the antisymmetric tensor.

 Finally, consider a group $G$ WZW model with level $k$,
which has a  $U(1)^d$
 global symmetry   ($\ie$ the background is independent of $d$
  coordinates). In order to obtain the $O(d)\times O(d)$
 duality one has to use an equivalent model
   $G\times U(1)^d/  U(1)_k^d$.  The WZW is
exact up to a shift $k\rightarrow k-\tilde c_G$, but
  the $O(d)\times O(d)$ duality  introduces the ${1\over k}$
corrections in the dual models.

\chapter{Summary}

In this paper we have derived a charged black hole solution
in four dimensions based on $SL(2,R)\times SU(2)\times U(1)/U(1)^2$
WZW coset model.
We
compared the semiclassical solution, obtained by integrating out the
gauge fields in the sigma model, to the exact to all orders solution
obtained by   the algebraic Hamiltonian approach.
We have seen that the space-time singularity exists  also in the
exact solution. Moreover, the structure of the
space-time described by
the exact metric depends strongly on the gauge parameter, unlike
in the semiclassical limit. According to the value of the gauge
parameter,
we have seen that a naked string
singularity or a surface (membrane) singularity could exist and the
black hole singularity can extends outside the horizon.
The exact vector dual model was derived explicitly as well.
In the semiclassical limit there is a naked singularity with
a topology $S^2\times R^1$. In the exact
 solution the space time structure  depends
on the gauge parameter, so that there might
appear  additional naked (string, membrane)
 singularities which do not show in the
semiclassical limit.

We have seen that the algebraic Hamiltonian approach is  useful
do study duality of metrics. In particular we were able to
determine how the exact metric and dilaton
transform under the $O(d,d)$
 duality and discovered that the $\O({1\over k})$ corrections
to the {\it inverse
metric}  (with respect to the semiclassical inverse metric)
are invariant under the $O(d)\times O(d)$ duality and only the
"semiclassical" part of it transforms. (The semiclassical part
transforms as in the one loop order transformations). Therefore,
although the algebraic Hamiltonian approach has a major disadvantage
of not being useful to calculate the antisymmetric tensor, knowing
the antisymmetric tensor to one loop order only is enough to obtain
the metrics and the dilaton
in all the {\it exact} $O(d,d)$  dual model.

 \appendix

This appendix contains the expressions for the Ricci tensor and
the scalar curvature that correspond to our charged black hole
solution.
In section 2 we have derived a semiclassical solution that corresponds
to the following metric (before the coordinate transformation on $r$
in \fifteen)
$$dS^2=-{\sinh^2r(1+Q^2\sin^2\theta)\over \cosh^2r
+Q^2\sin^2\theta}dt^2+
{k_1\over  k_2 }dr^2 + {\cosh^2 r\sin^2\theta\over \cosh^2 r
+Q^2\sin^2\theta}
d\phi^2+d\theta^2 \eqn\hifteen$$ with $Q^2=\tan^2\alpha$.
The corresponding exact metric
was derived in section 3 (again, before transforming $r$ in \uyg)
$$dS^2=-{\sinh^2r(1+C+Q^2\sin^2\theta)\over \cosh^2r
+Q^2\sin^2\theta+C}dt^2
+{(k_1-2)\over (k_2+2)}dr^2$$$$
+{\cosh^2\sin^2\theta\over \cosh^2 r+Q^2\sin^2\theta}d\phi^2
+d\theta^2\eqn\ueeg $$
with  $$Q^2={(k_2+2)k_1\over(k_1-2)k_2}\tan^2\alpha\eqn\qppq$$
and $$C={2\over k_1-2}(1-{k_1\over k_2}\tan^2\alpha)\eqn\cppc$$
We shall write both metrics as follows:
$$dS^2=-{\sinh^2r(B+Q^2\sin^2\theta)\over \Sigma_1}dt^2+adr^2
+{\cosh^2 r\sin^2\theta\over \Sigma_1}d\phi^2+d\theta^2\eqn\hgjb$$
where $\Sigma_1=\cosh^2 r+Q^2\sin^2\theta$ and $\Sigma_2=\cosh^2r
+Q^2\sin^2\theta+C$.
Eventually we can use the limit $C=0$ ($B=1$) and $a={k_1\over k_2}$
for the  semiclassical metric.
The Ricci tensor is the following:
$$R_{rr}=(B+Q^2\sin^2\theta)({-1\over \Sigma_2}+{3\cosh^2\over
\Sigma_2^2})+Q^2\sin^2\theta({-1\over \Sigma_1}+{3\sinh^2r\over
\Sigma_1^2})\eqn\rrrr$$
$$R_{tt}={\sinh^2 r\over \Sigma_2^2}({1\over a}(B+Q^2\sin^2\theta)(
{-3\cosh^2 r\over \Sigma_2}+{Q^2\sin^2\theta\over \Sigma_1}-1)$$$$
+Q^2\sinh^2r(\cos 2\theta-{\sinh^2r\sin^22\theta\over 4(B+
Q^2\sin^2\theta)}-{Q^2\sin^22\theta\over \Sigma_2}+{\cosh^2 r
\cos^2\theta\over \Sigma_1}))\eqn\rrtt$$
$$R_{\phi\phi}={\sin^2\theta\cosh^2r\over \Sigma_1^2}(
{4Q^2\over \Sigma_1}
({1\over a}\sin^2\theta\sinh^2r+\cosh^2r\cos^2\theta)
-{Q^2\over \Sigma_2}({1\over a}\sin^2\theta(B$$$$
+Q^2\sin^2\theta)
+{\cosh^2r\sinh^2r\cos^2\theta\over B+Q^2\sin^2\theta})+\cosh^2r(1+
Q^2-{1\over a}Q^2\sin^2\theta))\eqn\rrphi$$
$$R_{\theta\theta}={\cosh^2r\over \Sigma_1}+{3Q^2\cosh^2r\cos^2\theta
\over \Sigma_1^2}-{Q^2\sinh^2r\over (B+Q^2\sin^2\theta)^2\Sigma_2}
(B\cos 2\theta -Q^2\sin^2\theta)$$$$+{Q^4\sinh^2r\sin^22\theta\over
2(B+Q^2\sin^2\theta)\Sigma_2^2}(1-{1\over 2(B+Q^2\sin^2\theta})
\eqn\rrtheta$$
Finally, we give the expression for
the scalar curvature $R_{\mu}^{\;\mu}$. Here we use the coordinate
transformation $\cosh^2r\rightarrow r$
$$R_{\mu}^{\;\mu}={-Q^2{1\over a}\sin^2\theta+r(1+Q^2-{1\over a}Q^2
\sin^2\theta)\over \Sigma_1}
-{B+Q^2\sin^2\theta-1\over a\Sigma_2}$$$$
+{7Q^2({1\over a}(r-1)\sin^2\theta+r\cos^2\theta)\over \Sigma_1^2}
+{3{1\over a}r(B+Q^2\sin^2\theta+1)\over \Sigma_2^2}$$$$
-{Q^2(r-1)\cos 2\theta\over (B+Q^2\sin^2\theta)\Sigma_2}+
 {Q^2(r-1)({1\over 4}\sin^22\theta-B\cos 2\theta+Q^2\sin^2
\theta)\over (B+Q^2\sin^2\theta)^2\Sigma_2}$$$$
+{{3\over 2}Q^4(r-1)\sin^22
\theta\over (B+Q^2\sin^2\theta)\Sigma_2^2}
-{{1\over 4}Q^4(r-1)^2\sin^2 2\theta\over (B+Q^2\sin^2\theta)^2\Sigma_2
^2}$$$$-{Q^2\over \Sigma_1\Sigma_2}({1\over a}
\sin^2\theta(B+Q^2\sin^2\theta+1)+2
{r(r-1)\cos^2\theta\over B+Q^2\sin^2\theta})\eqn\curv$$
Carefully taking the limits, it is easily seen that the curvature
blows up  in three cases: (i) $r+Q^2\sin^2\theta=0$
(ii) $r+Q^2\sin^2\theta+C=0$ and (iii) $1+C+Q^2\sin^2\theta=0$.
In the semiclassical limit, since $C=0$ the only singularity is
at $r+Q^2\sin^2\theta=0$. However, in the exact solution $C$
can take any value. In particular, for
$ C\le -1$ there is an additional
singularity at  $\sin\theta= {-(C+1)\over Q}$.
For $C<-1$ this singularity is on cone surfaces and for $C=-1$ it
 is a string singularity.
\refout
\bye